\documentclass{ws-p8-50x6-00}

\begin{document}

\title{Ultra-high energy cosmic neutrinos}

\author{H. Athar}

\address{Department of Physics, Tokyo Metropolitan University, 1-1 
 Minami-Osawa, Hachioji-Shi, Tokyo 192-0397, Japan\\
 E-mail: athar@phys.metro-u.ac.jp}  

\maketitle

\abstracts{
 Several cosmologically distant astrophysical sources may produce ultra-high
 energy cosmic neutrinos ($E_{\nu} \geq 10^{6}$ GeV) of all flavors above the 
 atmospheric neutrino background. I study the effects of vacuum neutrino 
 flavor mixing on this cosmic neutrino flux. Prospects for observations of 
 these ultra-high energy cosmic neutrinos in large underwater/ice neutrino 
 telescopes are also briefly discussed.}

\section{Introduction}

	Search for ultra-high energy cosmic neutrinos 
 ($E_{\nu} \geq 10^{6}$ GeV)
will not only yield complemetary information about the highest energy 
 phenomena in the universe with respect to conventional very high energy 
 gamma ray 
astronomy but will also possibly address a fundamental aspect of particle 
 physics, that is, to corroborate the neutrino flavor mixing\cite{book}.

	This contribution is organized as follows: In section 2, I discuss in 
some detail the flux estimates, effects of vacuum neutrino flavor mixing and 
prospects
for observations of ultra-high energy cosmic neutrinos. In section 3, 
concluding remarks are given.

\section{Ultra-high energy cosmic neutrinos}

\subsection{Some flux estimates}

	A guarenteed source of ultra-high energy cosmic neutrinos is the 
interaction between ultra-high energy cosmic rays and the relic photon 
 flux\cite{relic}. Here, the flux for ultra-high energy cosmic (muon) 
 neutrinos, 
 $F^{0}_{\mu}$ is obtained by folding the observed ultra-high energy cosmic 
ray (power law) flux spectrum with the relic (thermal) photon flux spectrum 
 while 
integrating over the relevant differential cross-section and the kinematic
variables. This flux peaks  
typically at, $E_{\nu}\sim 10^{8}$ GeV with $F^{0}_{\mu}\sim 5\cdot 10^{-17}$
 (cm$^{2}$s sr)$^{-1}$. The position and height of the peak 
 depends on 
the assumed $z$ distribution of the sources, $f(z)$, for ultra-high energy 
 cosmic rays
as well as the distance traversed by these. This generic feature of peaking 
of ultra-high energy cosmic neutrino flux also holds for 
other possible sources such as Active Galactic Nuclei (AGN)
and Gamma Ray Burst fireballs (GRB), where the ultra-high energy cosmic rays 
 produced within the source interact with the ambient 
photon field present in the vicinity of source. The $E_{\nu}$ value at the 
 height of peak in $F^{0}_{\mu}$ depends on photon field spectrum shape. 
 The recent AMANDA and SuperK 
searches for such ultra-high energy cosmic neutrino flux provide useful
upper limits on this flux\cite{nu2000-1}.  

\subsection{Effects of vacuum neutrino flavor mixing}

	In the context of three flavors, the currently 
 preferred\cite{nu2000-2} flavor 
oscillation solutions to explain the solar electron neutrino deficit are the
LOW and the LMA (MSW), whereas for the atmospheric muon 
neutrino deficit, the currently favourable solution is the maximal depth 
flavor oscillations of $\nu_{\mu}$ into $\nu_{\tau}$. The present 
 observational status, thus provides indirect clues for a (quasi) bimaximal 
 neutrino mixing matrix. I will use the neutrino 
mixing parameters, ($\delta m^{2}, \, \sin^{2}2\theta$), corresponding to 
these  solutions to estimate the final 
 (downward going) ultra-high energy cosmic neutrino flux on earth, 
 $F_{\alpha}\, (\alpha = e, \mu, \tau)$ due to vacuum neutrino flavor mixing. 
 After averaging over the rapid 
 oscillations, the flavor precession probability is  
\begin{equation}
 \langle P(\nu_{\alpha} \rightarrow \nu_{\beta})\rangle \equiv 
 \langle P_{\alpha \beta}\rangle \simeq 
  \sum^{3}_{i=1}|U_{\alpha i}|^{2}|U_{\beta i}|^{2},
\end{equation}
where $U_{\alpha i}$ and $U_{\beta i}\, (\beta = e, \mu, \tau)$ stand for 
 the relevant elements of 3$\times $3 MNS 
 neutrino mixing matrix which I will use 
in standard parameterization. $F_{\alpha}$ in terms of 
 $\langle P_{\alpha \beta}\rangle $ and $F^{0}_{\alpha}$, the 
intrinsic ultra-high energy cosmic neutrino flux is 
\begin{equation}
 F_{\alpha}\equiv \sum_{\beta}\langle P_{\alpha \beta}\rangle  F^{0}_{\beta}.
\end{equation}
The $ \langle P_{\alpha \beta}\rangle $ matrix, in case of LOW 
solution for solar electron neutrino deficit alongwith the  
$\nu_{\mu}$ to $\nu_{\tau}$ flavor oscillations with maximal depth is
 
\begin{equation}
 \langle P_{\alpha \beta} \rangle = \left( \begin{array}{ccc}
                             1/2 & 1/4 & 1/4 \\
                             1/4 & 3/8 & 3/8 \\
                             1/4 & 3/8 & 3/8 
                            \end{array}
                      \right).
\end{equation}	
Using Eq. (3) in Eq. (2), it turns out 
that starting from $F^{0}_{e}: F^{0}_{\mu}: F^{0}_{\tau}$'s 
 as 1: 2: 0, one obtains $F$'s as 1: 1: 1 at the level of $F^{0}_{e}$, 
irrespective of the flavor oscillations solution to the solar electron 
neutrino deficit. Thus, for instance, the flavor composition in the guarenteed
flux for ultra-high energy cosmic neutrinos is expected to be equally 
distributed in $\nu_{e}, \, \nu_{\mu}$ and  $\nu_{\tau}$. 

If there is an incomplete or no averaging over the rapid oscillations, then
there will be $\delta m^{2}$ and $E_{\nu}$ dependences in 
 $P_{\alpha \beta}$ [see Eq. (1)]. In this
situation, the effects of $z$ distribution of the sources for 
ultra-high energy cosmic neutrinos should also be taken into account by 
 calculating $f(z)$ weighted $P_{\alpha \beta}$ where 
$P_{\alpha \beta}\, =\, \int^{z^{max}}_{0}P_{\alpha \beta}(z) f(z)dz/
\int^{z^{max}}_{0}f(z)dz$. It is relevant to  note that the matter enhanced 
flavor oscillation effects are negligible for the $\delta m^{2}$ and $E_{\nu}$ 
values under discussion.

\subsection{Prospects for observations}

	The typical km$^{2}$ effective surface area size neutrino telescopes 
 which
are currently under construction/planning may be able to search meaningfully
for ultra-high energy cosmic neutrinos. For instance, the guarenteed  
ultra-high energy cosmic neutrino flux gives several events
per kilometer per steradian in these neutrino telescopes. If the AGNs and GRBs
are also sources of ultra-high energy cosmic neutrinos then the event rate is
several orders of magnitude higher in the same neutrino telescopes depending 
 on the model of the AGN/GRB. Moreover,
in this case, even it is conceivable to identify the neutrino flavor content 
 in the ultra-high energy cosmic neutrino flux.

\section{Conclusion}

	For a comprehensive search of ultra-high energy cosmic neutrinos, (at
least) km$^{2}$ effective surface area size neutrino telescopes should have to 
be deployed.

\section*{Acknowledgments}
 This work is supported by a Japan Society for the Promotion of Science
 fellowship.

\end{document}